\documentclass[aps,prd,twocolumn,showpacs,preprintnumbers,amsmath,amssymb]{revtex4}
\usepackage{graphicx}
\usepackage{dcolumn}
\usepackage{bm}

\begin{document}
\title{Supernova Neutrinos: The Accretion Disk Scenario}
\author{G. C.~McLaughlin$^{1}$}
\author{R. Surman$^{2}$}
\affiliation{
$^{1}$ Department of Physics, North Carolina State University, \\
Raleigh, North Carolina 27695-8202. \\
$^{2}$ Department of Physics and Astronomy, Union College, 
Schenectady, NY 12308.\\}
  
\date{\today}

\begin{abstract}

Neutrinos from core collapse supernovae can be emitted from a rapidly 
accreting disk surrounding a black hole, instead of the canonical
proto-neutron star. For Galactic events, detector count rates are considerable
and in fact can be in the thousands for Super-Kamiokande.  The rate of 
occurrence of these accreting disks in the Galaxy 
is predicted to be on the order of 
$\sim 10^{-5} \, {\rm yr}^{-1}$, yet there is little observational 
evidence to provide an upper limit on their formation rate.
It would therefore be useful to
discriminate between neutrinos which have been produced in a 
proto-neutron star and those which have been produced accretion disks.
In order to distinguish between the two scenarios, either 
the time profile of the neutrino luminosity or the relative fluxes of
different neutrino 
flavors may be considered.   There are some signals that
would clearly point to one scenario or the other.

\end{abstract}
\maketitle

Recent advances in the theory of gamma ray bursts (GRBs) have pointed 
to rapidly 
accreting disks surrounding black holes (AD-BH) because this configuration
is a natural one for producing massive energy generation.  Simulations
of neutron star mergers and core collapse supernovae have indeed found that
under certain conditions, these disks can form 
\cite{MacFadyen:1998vz,Ruffert:1998qg,Rosswog:2005su}. 
Spectra from X-ray afterglows of long duration bursts 
have indicated that the bursts are
associated with Type Ib/Ic supernovae, suggesting massive stellar core
collapse as the origin.  The rate of gamma ray bursts in the galactic region
 is estimated to
be approximately $10^{-5} \, {\rm yr}^{-1}$ 
with approximately 2/3 of these being long duration events 
\cite{Podsiadlowski:2004mt}. This
is roughly 0.1\% 
of the core collapse supernova rate.
Disks which produce enough energy for a burst must have fairly high 
accretion rates in the range of 0.1 $M_\odot / \, {\rm s}$ to  
1.0 $M_\odot / \, {\rm s}$.   Observations of short bursts 
suggest that they are associated
with compact object mergers, neutron star - neutron star, or black hole-neutron
star \cite{hjorth}.

Hypernovae are very energetic core collapse supernova. 
The rate of Type Ic supernovae in an average galaxy is about $10^{-3} \, 
{\rm yr}^{-1}$, whereas the inferred hypernova 
rate is about $10^{-5} \, {\rm yr}^{-1}$
\cite{Podsiadlowski:2004mt} . A limited
data set of Type Ic hypernovae and supernovae shows that
supernovae which come from progenitors above $\sim 20 M_\odot$ show evidence
for  two branches with a few stars  
at high energy $\sim 10^{52}$ ergs, and one at
low energy $\sim 10^{50}$ ergs \cite{Nomoto}.  One can speculate
that the more luminous
events are produced by an accretion disk and the less luminous event is not.

It would be useful to translate these limits on hypernovae and 
gamma ray bursts into limits on the AD-BH formation rate.  Not every 
AD-BH may make a 
hypernova or GRB.  An observationally based 
upper limit can be made from
 the ratio of black holes to neutron stars. In the case of low-mass 
x-ray transients the observations will support turning
every star above $\sim 20 {\rm M}_\odot$ into a black 
hole \cite{Fryer:1999ht}, 
producing a rate of 10\% or even greater.  

\begin{figure*}
\vspace*{5.8cm}
\special{hscale=66 vscale=66 hsize=1500 vsize=600
         hoffset=0 voffset=0 angle=0 psfile="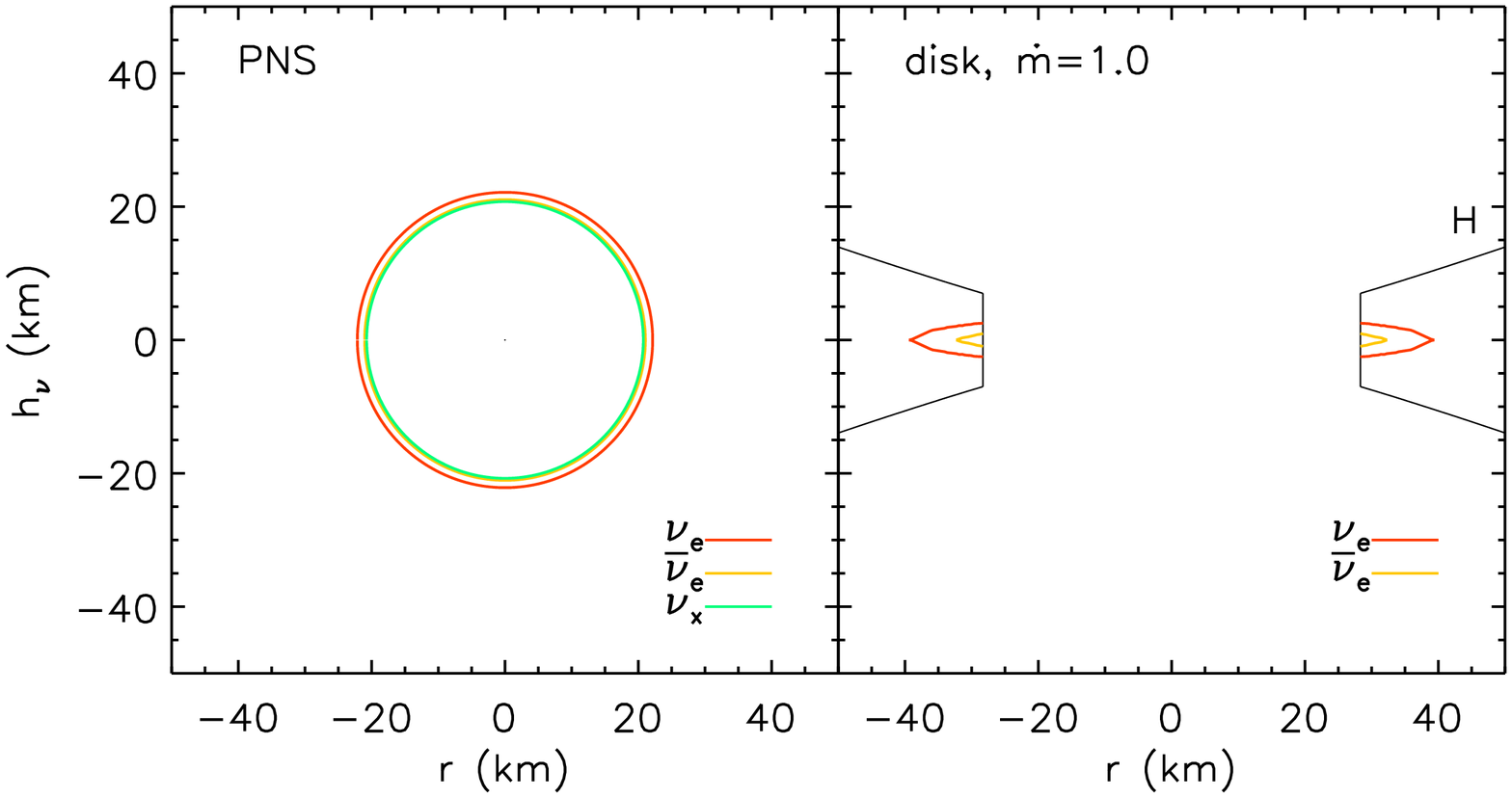"}
\caption{Shows the neutrino surfaces for a proto-neutron
star and a 1 $M_\odot / {\rm s}$ accretion disk.}
\label{fig:nuet_surf}
\end{figure*}

In this paper we point out that the next Galactic 
supernova may produce neutrinos in an accretion disk, not in a proto-neutron
star.  Even
if the supernova light curve is observed, it would be difficult
to use this to determine whether a PNS or an AD-BH is formed at the 
core.  In this paper we show that the emitted 
neutrinos may be used to immediately 
and directly make this distinction.  

Detectors that will observe neutrinos from the next galactic supernova are currently Super-Kamiokande \cite{SK}, KamLAND \cite{vogel}, MiniBooNE \cite{miniboone}, SNO \cite{Beacom:1998yb} and the future Borexino \cite{Cadonati:2000kq} .  A proto-neutron star event will produce 1000s of events in SuperK, and 100s in KamLAND and MiniBooNE.  At least one count from an AD-BH supernova may occur in SuperKamiokande even if the event is 3 Mpc away\cite{Nagataki:2000gy}. 

To estimate the neutrino signal, we first examine the disks from which they 
are emitted.  Neutrinos in accretion disks are produced primarily in inverse 
beta decay events. For sufficiently rapidly accreting disks the center of the disk has high enough temperature that the neutrinos become trapped. 
Fig. \ref{fig:nuet_surf} shows 
the surfaces 
of last scattering for the
neutrinos in the PNS and the AD. Fig \ref{fig:nuet_surf}b
is based on the calculations of \cite{Surman:2003qt}  for an
AD of $\dot{M} = 1.0 M_\odot / \, {\rm s}$, 
and black hole spin parameter a= 0,  
while Fig. \ref{fig:nuet_surf}a was estimated  
from the protoneutron star temperature and density profile of Ref. \cite{wilson}.  For a disk which accretes
at a rate 0.1 $M_\odot / \, {\rm s}$, the neutrinos are only barely
trapped.  

If one wishes to detect neutrinos from an AD-BH supernova, the total energy
emitted in the form of neutrinos is clearly important. For the
 PNS, analyses are often done by taking $3 \times 10^{53} \, {\rm erg}$
of gravitational binding energy liberated in the collapse and dividing it
equally between six species, three flavors of neutrinos and 
three of antineutrino.
In the case of the AD-BH, gravitational 
binding energy is released as material spirals in from the outer edges of the
disk but only some of that is released in the form of neutrinos.  The gravitational
binding energy released, given a 3 $M_\odot$ black hole, 
is $9 \, (M/M_\odot)  \times 10^{53} {\rm ergs}$,
where $M$ is the mass of material processed by the disk.  When
the neutrinos are not trapped the rate of energy loss is determined by the
rate at which neutrinos can be produced through inverse beta processes.
When neutrinos are trapped it is determined by the time it takes the neutrinos
to scatter out of the trapped region. We find that roughly 20\% of this 
gravitational binding energy is released in the form of neutrinos for disks 
of  1.0 $M_\odot  / \,  {\rm s}$ and 5\% in the case of 0.1 $M_\odot / \, {\rm s}$.

The timescale for neutrino emission through the proto-neutron star is 
about 10 s, with an exponentially decaying luminosity so that most of the 
neutrinos are emitted in the first three seconds.  If an accretion disk is 
steady state, the neutrino luminosity will not fall off but will be 
approximately constant. For example, for a 
0.01 $M_\odot \, / \, {\rm s}$ disk which processes 
1 $M_\odot$, the neutrino emission will be spread over 100 s,
although deviations from steady state will likely occur. 
The observed time profile of the neutrinos may
be used as a constraint.  

The energy spectra of the emitted neutrinos
will determine the flux averaged cross sections in supernova neutrino
detectors. Over  
the trapping surface in the $\dot{M} = 1 M_\odot / {\rm s}$ AD, the energies
are in the range of $\langle E_{\nu_e} \rangle \sim 13 \, {\rm MeV}$  to 
$\langle E_{\nu_e} \rangle \sim 14 \, {\rm MeV}$ and 
$\langle E_{\bar{\nu}_e} \rangle \sim 16 \, {\rm MeV}$ to 
$\langle E_{\bar{\nu}_e} \rangle  \sim 18 \, {\rm MeV}$ \cite{Surman:2003qt} .
  There is relatively 
little emission due to $\mu$ 
and $\tau$ type neutrinos in such disks although for faster accretion rates
such as 10 $M_\odot / s$,  $\mu$ and $\tau$ neutrinos are not only produced 
but also
become trapped.  In more slowly rotating, optically thin disks, the average energy of the neutrinos 
comes close to tracking
the average thermal energy of the electrons and positrons.  

\begin{figure*}
\vspace*{8.0cm}
\special{hscale=66 vscale=66 hsize=1500 vsize=600
         hoffset=0 voffset=0 angle=0 psfile="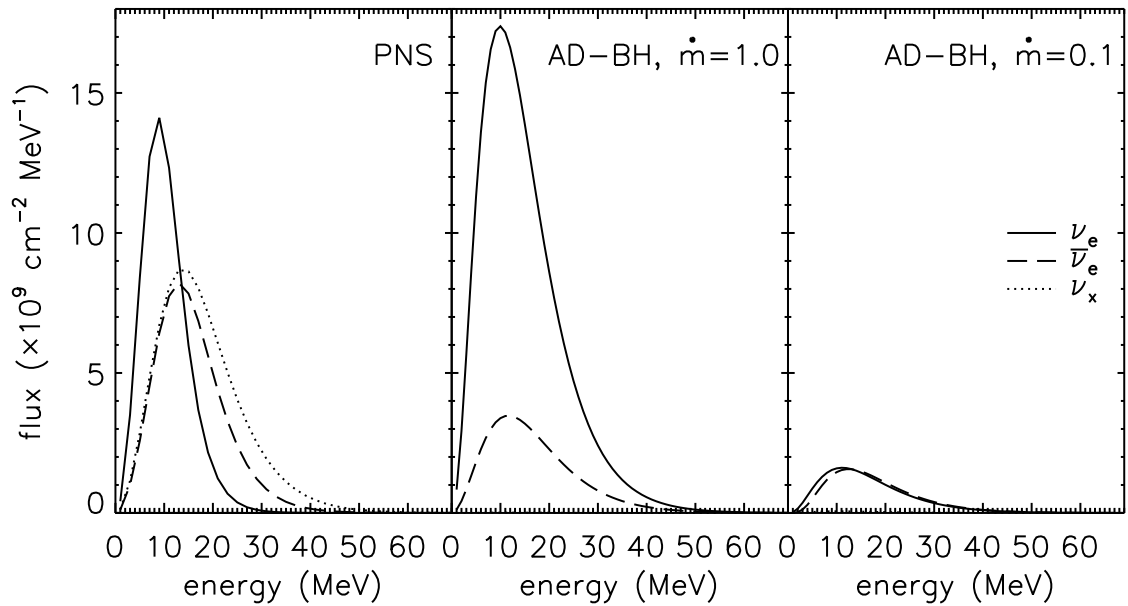"}
\caption{Shows the neutrino spectra for the PNS and AD-BH, in
the absence of oscillations, at a distance of 10 kpc from the center of the object.
The AD-BH neutrino spectra were calculated by summing over the entire surface of the disk
including both trapped and untrapped regions,
as described in  \cite{Surman:2003qt,Kneller:2004jr}.
The spectral parameters for the PNS
were taken from \cite{Keil:2002in}: 
$T_{\nu_e} = 2.6, \, \eta = 3.0 $,
$T_{\bar{\nu}_e} = 4.0, \, \eta = 2.8$, 
and $T_{\nu_\mu,\nu_\tau} = 5.0, \, \eta = 1.8 $, with an energy 
partition of $L_{\nu_e}:L_{\bar{\nu}_e}:L_{\nu_e}$ of $1:1.3:7.7$.}
\label{fig:spectra}
\end{figure*}

As can be seen from Fig. \ref{fig:spectra}, the spectra of the
$\nu_e$ and $\bar{\nu}_e$ are similar in the PNS and
AD-BH cases.  The muon and tau types however, are quite different,
suggesting that a detection of the neutrinos by flavor will give a
clear signature for one scenario or the other. However,
neutrino flavor transformation must be taken into account 
when translating the flavors of emitted spectra to the flavors that will 
be measured at the detector.

In both the PNS and AD-BH, the neutrinos
are emitted from a place of relatively high density, where they are matter 
suppressed and essentially coincident with matter eigenstates.
As the neutrinos 
proceed
to the outer edges of the star, which is at relatively low density, they
pass through two potential resonances regions and emerge as mass eigenstates,
see e.g. \cite{Dighe:1999bi}.
 The lower density resonance, 
at 
$\sim 10 \, {\rm g} \, {\rm cm}^{-3}$,
is governed by $\delta m_{12}$, and $\theta_{12}$ both of which have been measured
by solar and reactor neutrino experiments.  The higher density transition,
at $\sim 10^3 \, {\rm g}\,  {\rm cm}^{-3}$  is
governed by $\delta m_{13}$ and $\theta_{13}$ .  The former parameter
is known except for a sign (leaving the neutrino mass hierarchy unknown)
while the later is constrained by reactor neutrino experiments
$\sin^2 \theta_{13} < 0.1 $ \cite{Apollonio:2002gd}.  There is some
uncertainty regarding how the neutrinos will behave in these regions
due to the unknown hierarchy, $\theta_{13}$ and the density profile of
the star.

The results of various oscillation scenarios are shown in Table 
\ref{tab:fluxratios}, where we show the ratio of the total energy flux in all $\nu$s to that in $\nu_e$, $\bar{\nu}_e$ and the other four neutrinos as seen at 
a detector. A total energy flux could be inferred from a signal in the neutral
current channel, while the $\nu_e$ and $\bar{\nu}_e$ fluxes could be inferred
through measurements in the charged current channels.  It can be seen from
the table that if neutrinos
come from a PNS, then the energy flux ratios are relatively similar. 
The  ratios for the AD-BH cases, on the other
hand, vary considerably.
If $\nu_e$s or $\bar{\nu}_e$s encounter the higher resonance 
 adiabatically, then there is a clear signature 
for the AD-BH scenario because the flux ratios differ by an
order of magnitude from the PNS.  In almost all other cases, there are still differences
between the PNS and AD-BH flux ratios of a factor of two.

\begin{figure*}
\vspace*{5.8cm}
\special{hscale=66 vscale=66 hsize=1500 vsize=600
         hoffset=0 voffset=0 angle=0 psfile="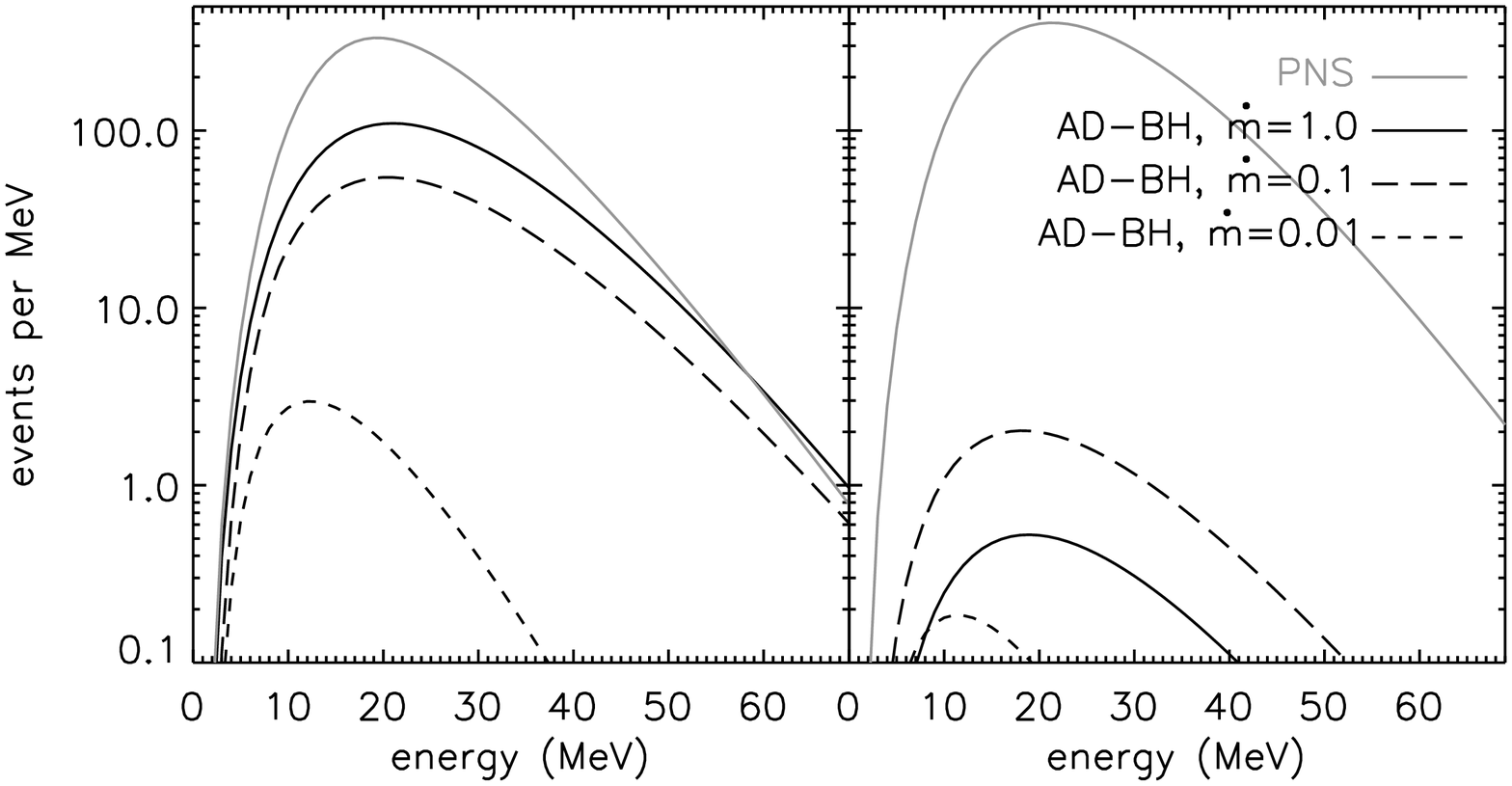"}
\caption{Comparison between the counts for ${\bar \nu}_e + p \rightarrow
e^+ + n$ at SuperKamiokande as a function of positron energy for the
two oscillation scenarios.  Plotted
are the signals from the proto-neutron star and  accretion disks of 
0.1 $M_\odot / \, {\rm s}$ to  1.0 $M_\odot / \, {\rm s}$.
With a normal hierarchy (left panel), 
a fraction of the neutrinos originally emitted
as $\bar{\nu}_e$s
are measured as $\bar{\nu}_\mu$ and $\bar{\nu}_\tau$s at the detector.
With an inverted hierarchy and an adiabatic $\theta_{13}$
resonance (right panel), almost all the neutrinos originally 
emitted as $\bar{\nu}_e$s
are measured as $\bar{\nu}_\mu$ and $\bar{\nu}_\tau$s at the detector. For the
PNS this makes relatively little difference as originally emitted
$\bar{\nu}_\mu$ and $\bar{\nu}_\tau$s are also converted to $\bar{\nu}_e$.
}
\label{fig:counts}
\end{figure*}

To determine the relative fluxes in the different channels, 
sufficient counts are required. In Fig. \ref{fig:counts} we
 compare events that would occur in SuperKamiokande, for a
supernova at 10 kpc, which collapses either to a PNS or to an AD-BH.
Estimates of total counts for
Super-Kamiokande, KamLAND,  
SNO (as a model for a $\nu_e$ sensitive detector) and OMNIS
 are given in Table \ref{tab:counts}.   The prospects for making the
distinction will depend on which detectors are on line at the time.  It
is essential to have a good neutral current signal measurement,
as this is how the total flux will be measured.  It is also vastly
preferable to have a way to measure both the $\nu_e$ and $\bar{\nu}_e$ flux if
one wishes to distinguish between a PNS and  an AD-BH.

\begin{table*}
\begin{tabular}{||c||ccc||ccc||ccc||ccc||ccc||} 
\hline 
& \multicolumn{3}{c||}{Normal Hierarchy}
& \multicolumn{3}{c||}{Normal Hierarchy}
& \multicolumn{3}{c||}{Normal Hierarchy}
& \multicolumn{3}{c||}{Inverted Hierarchy} 
& \multicolumn{3}{c||}{Inverted Hierarchy} \\
& \multicolumn{3}{c||}{H} 
& \multicolumn{3}{c||}{LA} 
& \multicolumn{3}{c||}{LNA} 
& \multicolumn{3}{c||}{H, LA} 
& \multicolumn{3}{c||}{H, LNA} \\ \hline
 &&&&&&&&&&&&&&& \\  
 & $ {\rm total} \over \nu_e$ & ${\rm total} \over \bar{\nu}_e$ & ${\rm total} \over 
\nu_x$ &  ${\rm total} \over \nu_e$ & ${\rm total} \over  \bar{\nu}_e$ & ${\rm total} \over \nu_x$ &  ${\rm total} \over \nu_e$ & ${\rm total} \over  \bar{\nu}_e$ & ${\rm total} \over \nu_x$ &  ${\rm total} \over \nu_e$ & ${\rm total} \over  \bar{\nu}_e$ & ${\rm total} \over \nu_x$ &  ${\rm total} \over \nu_e$ & ${\rm total} \over  \bar{\nu}_e$ & ${\rm total} \over \nu_x$  \\
&&&&&&&&&&&&&&& \\
 PNS, equipart  & 5.2 & 6.6 & 1.5 & 6.1 & 6.6 & 1.5 & 7.8 & 6.6 & 1.4 & 6.1 & 5.2 & 1.6 & 7.8 & 5.2 & 1.5 \\
 PNS, from \cite{Keil:2002in} & 6.0  & 6.0 & 1.5 & 6.0 & 6.0 & 1.5 & 6.0 & 6.0 & 1.5 & 6.0 & 6.0 & 1.5 & 6.0  & 6.0 & 1.5   \\
&&&&&&&&&&&&&&& \\
 $\dot{m} = 1.0$ from  \cite{Surman:2003qt}& 1600 & 7.6 & 1.2 & 4.0 & 7.6 & 1.6 & 1.8 & 7.6 & 3.2 & 4.0 & 1600 & 1.3 & 1.8 & 1600 & 2.3 \\
 $\dot{m} = 0.1$ from \cite{Surman:2003qt} & 78 & 3.0 & 1.5 & 6.5 & 3.0 & 2.0 & 3.0 & 3.0 & 3.0 & 6.5 & 78 & 1.2 & 3.0 & 78 & 1.5 \\
 $\dot{m} = 0.01$ from\cite{Surman:2003qt} & 41 & 2.7 & 1.7 & 7.6 & 2.7 & 2.0 & 3.8 & 2.7 & 2.8 & 7.6 & 41 & 1.2 & 3.8 & 41 & 1.4 \\
&&&&&&&&&&&&&&& \\
 \hline 
\end{tabular}
\caption{Ratios of energy fluxes for various accretion disk models and oscillation
scenarios as compared with the PNS. The PNS is calculated for the case of equipartition
of energy between neutrino species, and for the luminosity ratios ratios of Fig. \ref{fig:spectra}. 
  The quantity $\nu_x$ refers to the sum of
the $\nu_\mu$, $\nu_\tau$ and $\bar{\nu}_\mu$ and $\bar{\nu}_\tau$ luminosities.
Five oscillation scenarios are shown.  H means
that either neutrinos or antineutrinos, depending on the hierarchy, proceed adiabatically through the higher density
resonance (i.e. $P_{hop}=0$). 
 LA means that neutrinos proceed adiabatically through the lower density resonance, while
LNA means that neutrino pass nonadiabatically through the lower density resonance (i.e.
$P_{hop} = 1$).}
\label{tab:fluxratios}
\end{table*}

\begin{table*}
\begin{tabular}{|c||c|ccc||c|ccc||c|ccc||c|ccc||c|ccc| }\hline 
& & & $\dot{m}$ & 
&  & & $\dot{m}$ & 
&  & & $\dot{m}$ &  
&  & & $\dot{m}$ & 
&  & & $\dot{m}$ & 
\\
& PNS & $ 1.0$  &   $ 0.1$    &  $ 0.01$ 
& PNS & $ 1.0$  &   $ 0.1$    &  $ 0.01$ 
& PNS & $ 1.0$  &   $ 0.1$    &  $ 0.01$
& PNS & $ 1.0$  &   $ 0.1$    &  $ 0.01$ 
& PNS & $ 1.0$  &   $ 0.1$    &  $ 0.01$
\\\hline
SuperK 
&      &    &      &     
     &    &    &     & 
     &      &    &      &   
     &      &    &      &  
     &      &    &      & \\
$\bar{\nu}_e + p \rightarrow n + e^+$ 
     & 7000 & 2800  & 1400 & 50 
     & 7000 & 2800 &  1400 & 50 
     & 7000 & 2800 &  1400 & 50    
     & 10000 & 10 & 50  & 2 
     & 10000 & 10   & 50     & 2  \\
$\nu + ^{16}{\rm 0} $ 
     & 30  & 20  &  3.0    & 0.3 
     & 30  & 20  &  3.0    & 0.3 
     & 30  & 20  &  3.0    & 0.3   
     & 30  & 20  &  3.0    & 0.3 
     & 30  & 20  &  3.0    & 0.3   \\
SNO &      &    &      &     
     &    &    &     & 
     &      &    &      &   
     &      &    &      &  
     &      &    &      & \\
$\nu_e + d \rightarrow p + p + e^-$ 
     & 170     &  0.2    &  0.8     & 0.04    
     & 130   &    80   &    10   & 0.2 
     & 80     &  180   &  20     & 0.4  
     & 130     &  80   &  10    &  0.2
     & 80     &  180   &  20    &  0.4 \\
 $\bar{\nu}_e + d \rightarrow n + n + e^+ $ 
     & 70   &  30  & 0.2    & 0.4     
     & 70   &  30  & 15    & 0.4
     & 70   &  30  & 6    & 0.4  
     & 100  &  0.1  & 0.5  & 0.02 
     & 100  &  0.1  & 0.5  & 0.02  \\
 $\nu + d \rightarrow p + n + \nu$ 
     & 330  & 140  & 30     & 0.7    
     & 330  & 140  & 30    &  0.7
     & 330  & 140  & 30     & 0.7  
     & 330  & 140  & 30     & 0.7 
     & 330  & 140  & 30     & 0.7  \\
KamLAND      &      &    &      &     
     &    &    &     & 
     &      &    &      &   
     &      &    &      &  
     &      &    &      & \\
$\bar{\nu}_e + p \rightarrow  n + e^+$ 
     &  280    & 110   &  50    & 2    
     &  280    & 110   &  50    & 2 
     &  280    & 110   &  50    & 2  
     &  380    & 0.5   &  3    & 0.1  
     &  380    & 0.5   &  3    & 0.1  \\
$\nu + p \rightarrow \nu + p$ 
     &  500    &  200   &  44    &  1   
     &  500    &  200   &  44    &  1
     &  500    &  200   &  44    &  1  
     &  500    &  200   &  44    &  1 
     &  500    &  200   &  44    &  1  \\
OMNIS 
     &      &    &      &     &    &    &     & 
     &      &    &      &   
     &      &    &      &  
     &      &    &      &   \\
$\nu_e + {\rm Pb} \rightarrow e^- + n + {\rm Bi}$ 
     &  210    &  0.1    &  0.9    & 0.2     
     &  150    &  90     &  10     &  0.1
     &  80     &  190    &  30     & 0.3  
     &  150    &  90     &  10     &  0.1
     &  80    &  190     &   30    &  0.2  \\
 $\nu +  {\rm Pb} \rightarrow \nu + {\rm Pb} + n$ 
     & 40     & 16   &  4    & 0.04  
     & 40     & 16   &  4    & 0.04
     & 40     & 16   &  4    & 0.04  
     & 40     & 16   &  4    & 0.04 
     & 40     & 16   &  4    & 0.04  \\
\hline
\end{tabular}
\caption{Integrated counts in various detectors. The columns represent the
same oscillation
scenarios as in Table \ref{tab:fluxratios}. All accretion disk rates may be scaled by $(M/M_\odot)$ 
where M is the mass of material processed by the disk. OMNIS rates are quoted for 1kt of lead perchlorate only.
More detailed discussion of count rates for the PNS are in, e.g. 
\cite{Langanke:1995he,Beacom:2002hs,Boyd:2003em, Beacom:1998yb}.}
\label{tab:counts}
\end{table*}

We have discussed the observational consequences of the next 
Galactic core collapse
event forming at its center an accretion disk surrounding
a black hole instead of a proto-neutron star.  This type of event will
produce significant event rates in currently on-line neutrino detectors, with
a time profile that will likely differ from that of the canonical PNS. 
With a neutral current measurement,  
it may be possible to determine the origin of core-collapse supernova 
neutrinos 
by inferring the ratio of the total energy flux  
 to charged current flux.  For the AD-BH scenario this could be as large 1600:1.

This work was supported in part  by the Department of Energy, 
under contract DE-FG02-02ER41216 (GCM), DE-FG05-05ER41398 (RS).
and by the National Science Foundation under Grant No.
PHY99-7949.

\end{document}